\documentclass[11pt,twoside]{article}  
\usepackage{asp2006}
\usepackage{adassconf}

\begin{document}   

%

\paperID{P8.8}

%

\title{GOSSIP, a new VO compliant tool for SED fitting}
       
%
%
%
%
%

\markboth{Franzetti et al.}{GOSSIP, a new VO compliant tool for SED fitting}

%

\author{P. Franzetti, M. Scodeggio, B.\ Garilli, M. Fumana, L.\ Paioro}
\affil{INAF - IASF Milano, Milan, Italy}






\contact{Paolo Franzetti}
\email{paolo@lambrate.inaf.it}

%
%

\paindex{Franzetti, P.}
\aindex{Scodeggio, M.}
\aindex{Garilli, B.}
\aindex{Fumana, M.}
\aindex{Paioro, L.}





\begin{abstract}          
We present GOSSIP (Galaxy Observed-Simulated SED Interactive Program), 
a new tool developed to perform SED fitting in a simple, user friendly 
and efficient way.
GOSSIP automatically builds-up the observed SED of an object (or a large 
sample of objects) combining magnitudes in different bands and eventually 
a spectrum; then it performs a $\chi^2$ minimization fitting procedure 
versus a set of synthetic models. The fitting results are used to estimate
a number of physical parameters like the Star Formation History, 
absolute magnitudes, stellar mass and their Probability Distribution Functions. 
User defined models can be used, but GOSSIP is also able to load models produced 
by the most commonly used synthesis population codes.
GOSSIP can be used interactively with other visualization tools using the PLASTIC 
protocol for communications.
Moreover, since it has been developed with large data sets applications in mind, 
it will be extended to operate within the Virtual Observatory framework.
GOSSIP is distributed to the astronomical community from the PANDORA group
web site (\makeURL{http://cosmos.iasf-milano.inaf.it/pandora/gossip.html}).
\end{abstract}


\section{Introduction}

GOSSIP is a tool created to fit the electro-magnetic emission of 
an object (the SED, Spectral Energy Distribution) against synthetic models, to find
the simulated one that best reproduces the observed data. It has been developed to perform 
this task in a simple, user friendly and efficient way.\\
GOSSIP was born within the frameworks of the VVDS (Le F\'evre et al. 2005) and the zCOSMOS (Lilly et al. 2007) 
surveys; therefore it has been optimized to work on huge amounts of data as the ones provided by modern 
photometric and spectroscopic surveys.\\
GOSSIP has been written by the \htmladdnormallinkfoot{PANDORA Group}{http://cosmos.iasf-milano.inaf.it/pandora} at 
INAF IASF-Milano, using the \htmladdnormallinkfoot{Python}{http://www.python.org} language for 
the graphical part and the C language for the high performance computational tasks.

\section{GOSSIP flow chart}

\subsection{SED building}
\begin{figure}[t]
\epsscale{1.0}
\plotone{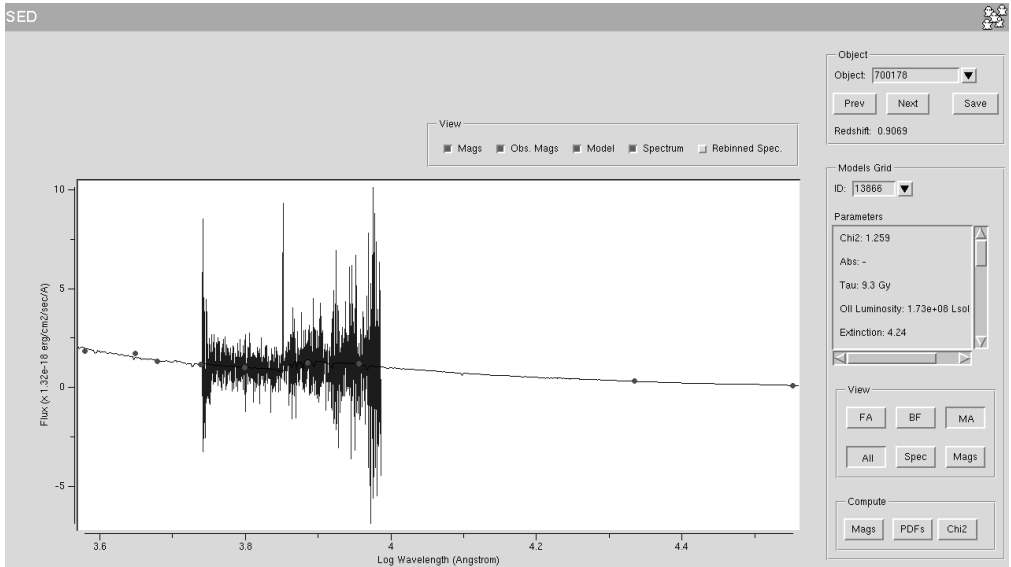}
\caption{The GOSSIP SED visualization window. In this example it is shown a SED composed by an optical spectrum (the rapidly
varying thick solid line) and magnitude points (the dots). This zCOSMOS SED is composed by data from VLT-VIMOS, CFHT, SUBARU
and SPITZER data. The best-fitting PEGASE model (thin smooth line) is over-imposed to the data while its main parameters are 
summarized in the right panel.} \label{sed}
\end{figure}

GOSSIP builds-up the observed SED of an object combining magnitudes in different bands from different 
instruments and, eventually, also a spectrum. Data can be read both from ASCII files or from a MySQL database
table. Figure \ref{sed} shows an example of a SED built by GOSSIP using data from the zCOSMOS survey.

\subsection{Synthetic models}

\begin{figure}[t]
\epsscale{0.7}
\plotone{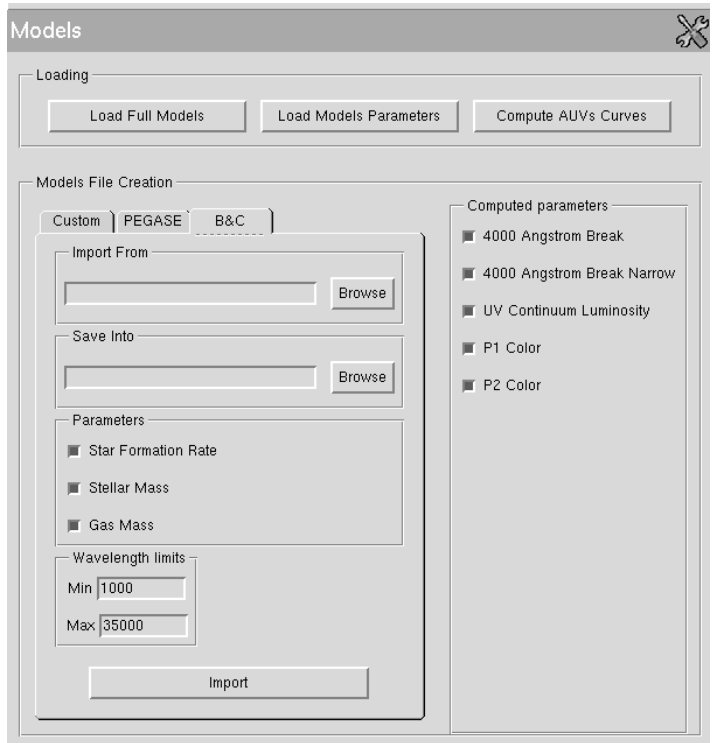}
\caption{The Bruzual \& Charlot models import panel. The user can choose original models parameters to be loaded and 
also parameters to be computed by GOSSIP itself.} \label{models}
\end{figure}

GOSSIP fits sets of synthetic models against observational data. It can use user-defined models, or ``standard'' 
ones computed by the most commonly used synthesis population codes like PEGASE (Fioc \& Rocca-Volmerange 1997) or 
Bruzual \& Charlot (Bruzual \& Charlot 2003) by directly reading the output files produced by these codes. It can also 
compute a set of parameters on the loaded models (both user-defined and ``standard'') like a photometric color or a
spectral index. In figure \ref{models} the Bruzual \& Charlot import panel is shown.

\subsection{The fitting procedure}

GOSSIP performs a $\chi^2$ minimization to find the model which best represents observational data in
interactive or in batch mode. \\
The minimization results for each object are stored into a file, so that different filters can be applied to them
to select the best-fitting model. Model ages can be left unconstrained, or bounded by the age of the Universe at the 
object's redshift, or the user can select a fixed galaxy formation epoch for the whole sample. Moreover the input
data set can be limited to the use of only the photometric or the spectroscopic data.  
Fitting results can be plotted in the main SED visualization window (see figure \ref{sed}) where the user can switch 
very rapidly between the various filters. \\
The fitting procedure can run on a single CPU or on a BEOWULF cluster; porting to the computational GRID is currently on-going. 
Once the fitting procedure has been performed, a number of ``post-fitting'' operations are carried out, which include the computation 
of physical parameters, like absolute magnitudes and stellar masses, from the best-fitting model and
of their full Probability Distribution Functions from the full set of synthetic models.

\subsection{Interoperability}

GOSSIP can connect to a running \htmladdnormallinkfoot{PLASTIC}{http://plastic.sourceforge.net/} hub to send the fitting results to other 
specialized tools, like TOPCAT or VISIVO for visualization.

\section{GOSSIP and the Virtual Observatory}

A \emph{standard} VO interface within GOSSIP is currently under development. Using this interface GOSSIP will be able 
to download single spectra from a SSA service to be fitted against synthetic models. However, since a relatively narrow   
wavelength coverage results into large uncertainties in the SED fitting results, a single spectrum very poorly constrains
a GOSSIP fit. The results can be significantly improved using also magnitudes to extend the wavelength coverage.
Individual aperture magnitude measurements in several optical and NIR filters can be currently gathered using VO protocols, but
they are not suited for a correct SED fitting, as detailed information about the aperture at which each data 
point has been obtained is required to build up a reliable SED. Aperture characterization is already provided 
in the Spectrum Data Model (McDowell, 2007), but it is not yet much used by data providers.
An alternative approach could be to download a ready-made observed SED provided by some VO service;
in this case the SED would be built by the data provider after having properly normalized the magnitude measurements
to a single common aperture. Unfortunately the SED Data Model is still under definition and it is not yet possible to 
obtain ready made SED through the VO. 
Another limitation we are facing, irrespective of the approach used to build a SED, is the lack of a detailed characterization of 
the magnitudes. The SED fitting procedure requires the knowledge of the exact shape of the filter transmission curve associated to 
each magnitude in order to compute the synthetic magnitudes from the models that are used in the $\chi^2$ minimization. However
a specification of the filter response curve is not yet included in the Spectrum Data Model.
Despite the fact that these limitations prevent SED fitting within the current VO implementation, GOSSIP is being developed 
paying close attention to the development of the VO technology. When these issues will be solved, it will be ready
to exploit all the VO potentialities.

\section{Conclusions}

We have presented GOSSIP, a new tool dedicated to SED fitting. It has been developed in order to perform this task 
in a simple, user friendly and efficient way. 
GOSSIP builds-up observed SEDs, loads synthetic models, performs $\chi^2$ minimization fitting between them and estimates
physical parameters for huge samples of objects.
The features implemented within this program make it a very useful tool for the analysis of the large data samples currently
available to the astronomical community.



\begin{references}

\reference Bruzual, G. \& Charlot, S. 2003, MNRAS, 334,1000
\reference Fioc, M. \& Rocca-Volmerange, B. 1997, A\&A, 326, 950
\reference Le F\'evre, O. et al.\ 2005, A\&A, 439,845
\reference Lilly, S. et al.\ 2007, ApJS, 172, 70
\reference McDowell, J. et al. 2007, IVOA Documents, Spectrum Data Model Version 1.02,\\
available at: \makeURL{http://www.ivoa.net/Documents/latest/SpectrumDM.html}
    
\end{references}
\end{document}